\documentclass[twocolumn,showpacs,amsfonts]{revtex4-1}

\usepackage{amsmath}
\usepackage{latexsym}
\usepackage{float}
 \usepackage{amssymb}
\usepackage{graphicx}
\usepackage{textcomp}
\usepackage{hyperref}
\usepackage{subfigure}

\def\ba{\begin{eqnarray}}
\def\ea{\end{eqnarray}}
\def\be{\begin{equation}}
\def\ee{\end{equation}}
\def\bm{\begin{math}}
\def\me{\end{math}}

\newcommand{\dummy}

\begin{document}
\title{ 
Finite-size behavior in phase transitions and scaling in the progress of an epidemic
}
\author{Subir K. Das$^{*}$}
\affiliation{Theoretical Sciences Unit and School of Advanced Materials, 
Jawaharlal Nehru Centre for Advanced Scientific Research,
 Jakkur P.O, Bangalore 560064, India}

\date{\today}
\begin{abstract}
	Analytical descriptions of patterns concerning spread and fatality
	during an epidemic, 
	covering natural as well as restriction periods, are important for reducing damage.
	We employ a scaling model to investigate this aspect 
	in the real data of COVID-19.
	It transpires from our study that the statistics of fatality 
	in many geographical regions follow unique behavior,
	implying the existence of universality. Via Monte Carlo simulations in finite boxes, we also present results
	on coarsening dynamics in a generic model that is used frequently for the studies of 
	phase transitions in various condensed matter systems. 
	It is shown that the overall growth, in a large set of regions, concerning disease dynamics and 
	that during {\it phase transitions} in a broad class of {\it finite} systems are closely related to each other and
	can be described via 
	a single and compact mathematical function, thereby providing a new angle to the studies of the former.
	In the case of disease we have demonstrated that the employed scaling model has predictive capability,
	even for periods with complex social restrictions.
	These results are of much general epidemiological relevance.
\end{abstract}


\maketitle
\section{Introduction}

COVID-19, a disease spread by a novel coronavirus, has produced 
global harm of historic magnitude \cite{wiki}. Many countries suffered
severely from multiple waves of infections \cite{wiki}. 
Despite social restrictions (SR),
a large group of countries fought the first wave for many months \cite{wiki}. Naturally, the topic of
epidemic is of much recent research interest \cite{mug, tian, cesar, lamm, skd_spread, adam, fer, ma}.
For combating communicable diseases, via accurate predictions, 
there exists longstanding interest in finding appropriate mathematical pictures \cite{kerm, ander,hjh}.
It is important 
to show that the growth in spread, as well as in fatality,  
combining the early natural and the late time SR 
periods of a wave, within an epidemic, can be accurately described by a compact mathematical form. Furthermore,
for a general understanding of universality in nonequilibrium systems, and thus,
for benefiting from possible existing information,
it should be instructive to compare the scenario of disease dynamics with growth phenomena in   
other systems, e.g., in systems undergoing 
phase transitions \cite{fisher1,onuki,bray,fisher3,book1,df,binder2,majumder,majumder2,das0,majumder3,majumder4}.
Outcomes of such studies, in addition to being of fundamental importance, can have
practical relevance in the containment of damage due to the spread of a disease.

An expectation for the early period natural spread of an 
infectious disease is the exponential rise 
\cite{kerm, ander, siam, stanley, mug, tian, smm, cesar, lamm, skd_spread}
\begin{equation}\label{gexp}
	N_i=N_{0i}\exp(m_i t),
\end{equation}
where $N_i$ is the number of infections till time $t$, with $N_{0i}$ and $m_i$ being constants.
Such a behavior may be present in the fatality rate as well. 
A deviation from Eq. (\ref{gexp}) should occur at later time. This
can be due to natural reasons as well as because of SR
\cite{skd_spread,
kerm, ander, siam, stanley, mug, tian, smm, cesar, lamm, adam,fer,ma}.
The post-exponential behavior can have different characters. Appearances of peaks in daily new cases, while
the total number of infections remaining much smaller than the population of a country, 
suggest that if strict measures are put in place,
economic scenario permitting and human patience persisting, the epidemic may die soon. 
The problem, in general, is of {\it quite} finite-size type 
\cite{fisher3,book1,df,binder2,majumder,majumder2,das0,majumder3,majumder4}, if
multiple infections of a single individual is disregarded. 
Thus, we use a finite-size scaling approach (from the literature of phase transitions) to describe the phenomena
\cite{skd_spread,fisher3,book1,df,binder2,majumder,majumder2,majumder3,
majumder4,das0}.
However, for disease dynamics the reason behind finiteness is not unique and 
infected population, at the end of an epidemic, can be a rather small fraction of the overall inhabitants in a given region,
unlike the standard phase transition behavior \cite{fisher3,book1,df,binder2,majumder,majumder2,das0,majumder3,majumder4} 
of materials (of finite size) that are studied in computers. 
This has connection with SR, acquired herd immunity, mutations of the pathogens, etc.

The applied approach reveals that the growth covering a {\it complete} 
wave for COVID-19, in a large set of geographical regions, can be described by a
single analytical function, implying the existence of universality. We have demonstrated that this
mathematical formulation provides predictive character to the model.
We also present results from the computer simulations \cite{book1}
of coarsening dynamics during {\it phase transitions} \cite{fisher3,book1} in 
systems, having finite sizes, described by a well-known model Hamiltonian. The above mentioned function,
interestingly, describes the latter phenomena also rather well. For disease, even though we discuss both
spread and fatality, the focus here is on the latter.

\section{Models and Methods}

Average length of domains ($L$), rich in one or the other species, e.g., A and B particles in a (A+B) binary mixture, 
during phase separation grows with time
typically as \cite{book1, bray, majumder} $L = L_0 + at^{\alpha}$. Here $L_0$ is the value of $L$
at $t=0$ and $a$ is an amplitude. In finite systems, having $N$ particles, the above mentioned
time dependence is obeyed only at early times. At late enough times, one has $L=L_{\rm max}$, 
with $L_{\rm max}$, the saturation value, being proportional to $N^{1/d}$ in $d$ dimensions. These two
limiting facts are bridged via the introduction of a scaling function $Y(y)$ such that \cite{binder2, majumder, majumder2}
\begin{equation}\label{scl1}
(L-L_0)^{1/\alpha}=Y(y)(L_{\rm max}-L_0)^{1/\alpha},
\end{equation}
where $y$ ($=L_{\rm max}^{1/\alpha}/t$) is a dimensionless scaling variable. 
$Y$ is independent of system size, i.e., for optimum choice of $\alpha$ there should be
collapse of data from different system sizes. This fact is exploited in this {\it finite-size scaling} method to
quantify the growth in a system in the thermodynamically large size limit.
When $y\rightarrow \infty$, one expects \cite{binder2, majumder, majumder2} 
$Y\sim y^{-1}$ so that early time behavior is recovered. For $y\rightarrow 0$, i.e.,
in the limit $t\rightarrow \infty$, $Y$ should converge to a constant \cite{binder2, majumder, majumder2}.

For the study of the above facts during 
phase separation we have performed Kawasaki exchange Monte Carlo (MC) simulations \cite{book1} of the 
nearest-neighbor
ferromagnetic Ising model \cite{fisher1, book1}, on 2D square lattices, set inside
square boxes of different sizes. The model is 
described by the Hamiltonian $H=-J\sum_{<ij>} S_i S_j$, with interaction strength $J>0$ and $S_i=\pm 1$, different signs 
of the spin variable corresponding either to an A or a B particle.
The coarsening dynamics in this model was studied by 
quenching infinite temperature homogeneous initial configurations to a temperature $T=0.6T_c$,
where $T_c$ ($\simeq 2.27$) is the critical temperature \cite{book1}, in units of $J/k_B$, $k_B$ being
the Boltzmann constant.
A trial move in our simulations is an exchange between a randomly chosen pair of nearest neighbor
spins \cite{book1}. These moves are accepted according to the standard Metropolis algorithm \cite{book1}.
$N$ such trial moves make a MC step (MCS), the unit of time \cite{book1}. Lengths from the simulation
snapshots were obtained by exploiting a well known scaling property of the two-point 
equal time correlation function \cite{bray} that complies with the self-similarity feature of growing domains \cite{bray}.
We have applied periodic boundary conditions and worked with $50$:$50$ compositions of
+ve and -ve spins. The presented simulation results were
averaged over about $50$ independent initial configurations.
For this model, one has \cite{lif,majumder,majumder2,das0,majumder4,huse,amar} 
$\alpha=1/3$. 

In analogy with the finite-size scaling model of kinetics of
phase transitions \cite{fisher3,book1,binder2,majumder,majumder2}, we briefly discuss here an analytical construction
for the disease dynamics, following Ref. \cite{skd_spread}. For the latter, the finite-size limit of a system, in the case
of fatality, is $N_{sf}$, the total number of deaths when the epidemic stops, which
we take {\it ideally} as the perfect end of the first wave. Considering that the early time fatality may follow Eq. (\ref{gexp}) as well, like the spread,
the (finite-size) scaling ansatz can be written as 
\begin{equation}\label{scl2}
	\ln n_f = Y(y) \ln n_{sf},
\end{equation}
with $y=\ln n_{sf}/ m_f t$, $n_f = N_f / N_{0f}$ and $n_{sf} = N_{sf}/N_{0f}$.
Here $m_f$ is the pre-factor of time inside the exponential, $N_{0f}$ is the fatality at the onset 
of exponential behavior and $N_f$ is the fatality at time $t$ since the onset of the
exponential behavior. Note that logarithmic conversions in this case are by anticipating exponential
early growth. For simple exponential growth \cite{skd_spread}, like in Eq. (\ref{gexp}), 
one expects $Y\sim y^{-1}$, for $y \gg 0$ and $Y$ should approach
a constant for $y\rightarrow 0$. Given that $N_{sf}$ is unknown, when the epidemic is still in progress, in the following
we will work with $N_{df}$. The latter is the number at which deviation of the time dependence 
of fatality occurs from the rise in accordance with the {\it thermodynamic limit} behavior, here {\it mostly} complying with the 
exponential growth. 
We will exploit the fact that $N_{df}\propto N_{sf}$,
which we know to be true from the experience with phase transitions. We will use $N_{df}$ as an
adjustable parameter, along with $N_{0f}$, for optimum data collapse.
Later, for the comparison with
the scaling plot from phase transitions, the ordinate for one of the master functions has to be adjusted
via a multiplication factor $Y(y=0)$, in addition to horizontal shifting by
another constant factor that is supposed to take care of the simple {\it time mismatch} between the two cases.

Starting from an instantaneous exponent \cite{majumder,huse,amar} ($\psi_i = d\ln Y/d\ln y$) approach 
(see Ref. \cite{skd_spread} for details), 
by accepting that for large $y$ one has power-law
character of $Y$ with exponent $p$ ($p$ being unity for simple exponential growth of $N_f$), one may write
\begin{equation}\label{scl3}
	-\frac{1}{\psi_i}=\frac{1}{p} + f(y).
\end{equation}
Here $f(y)$ is the deciding function for the convergence of $Y$ to a constant value, in the $y\rightarrow 0$ limit,
following deviation from the power law. 
For $f(y)=b y^{-\theta}$, one obtains \cite{skd_spread}
\begin{equation}\label{scl4}
	Y(y)=Y_0 \left(b+\frac{y^{\theta}}{p}\right)^{-p/\theta},
\end{equation}
where $Y_0$, $b$, $p$, and $\theta$ are positive constants, with $Y(y=0)=Y_0/b^{p/\theta}$. 
Combining Eqs. (\ref{scl2}) and (\ref{scl4}),
fatality as a function of time can be obtained. Note here that the unit of time in the disease dynamics is a day.

\section{Results}

First we provide a discussion on the disease dynamics.
Results are presented for whole of Europe as well as five countries within it, viz., Germany (DE),
Netherlands (NL), Spain (ES), United Kingdom (UK), and Italy (IT). The first infections in these
places were reported on the following dates \cite{wiki} of 2020: Europe -- 27 January (Jan); DE -- 27 Jan; NL -- 27 February (Feb); 
ES -- 31 Jan; UK -- 31 Jan; IT -- 31 Jan. The first fatalities were reported on \cite{wiki}:
Europe -- 15 Feb; DE -- 9 March; NL -- 6 March; ES -- 8 March; UK -- 5 March; IT -- 21 Feb, again all in 2020.
In each of the cases we have worked with data for at most up to first $120$ days. We have analyzed data from more places,
but confined ourselves, with respect to presentation, within a set that form a universality class in certain strict sense.

\begin{figure}[h!]
\centering
\includegraphics*[width=0.4\textwidth]{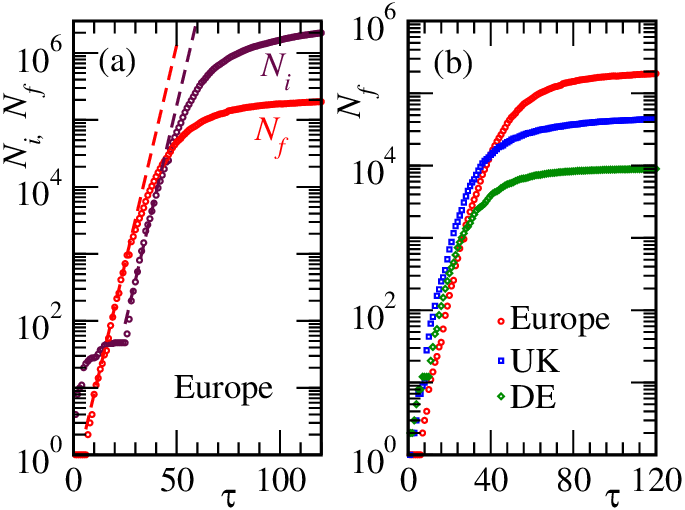}
        \caption{\label{fig1} (a) Plots of total number of infections ($N_i$) and
        fatalities ($N_f$) in Europe, with the variation of time. A semi-log scale has been chosen.
        The dashed lines correspond to exponential behavior.
        (b) Time dependence of fatalities in Europe and two countries within the continent
        are shown in a semi-log scale.}
\end{figure}

In Fig. \ref{fig1} (a) we show, versus time $\tau$, the total numbers of infections and deaths in Europe.
This time was counted from the day on which the first cases in respective categories were reported.
In both the plots exponential enhancements are visible for significant
periods of time, after an onset time $\tau_0$. This can be appreciated from the agreements of the data sets with the dashed lines
that represent simple exponential growths as in Eq. (\ref{gexp}). 
Fig. \ref{fig1} (b) shows $N_f$, versus $\tau$, from Europe and two countries inside the continent,
viz., UK and DE. In the rest of the manuscript all the
COVID-19 results will be presented on fatality only.

\begin{figure}[h!]
\centering
\includegraphics*[width=0.4\textwidth]{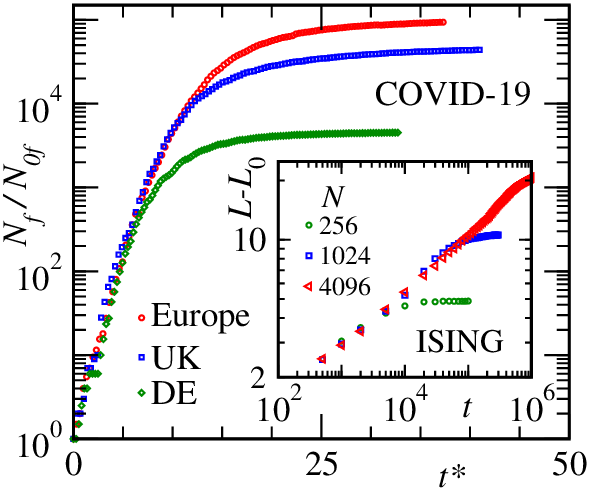}
        \caption{\label{fig2} $N_f/N_{0f}$, for Europe, DE and UK,
        are shown with the increase of time, on a semi-log scale. The time here,
	for each of the regions, has been translated with respect
        to the value at which a jump, signifying the onset of exponential behavior, was observed.
        Furthermore, this translated time was scaled via multiplication by appropriately chosen
        $m_f$, value of which is region specific. Inset: Plot of growth data for the Ising model,
        from different system sizes, on a log-log scale. We have used $L_0=1.29$, a value that
        emerges from the simulations.}
\end{figure}

In Fig. \ref{fig2} we show plots of $N_f/N_{0f}$, versus translated and scaled time
$t^*$ [$=m_f(\tau-\tau_0)$], on a semi-log scale. After such transformations early data from different places collapse 
on top of each other. At late time there is deviation, of course, which is related to the departure
from the exponential behavior. The overall picture is analogous to that of kinetics of {\it phase transitions} in finite box of different sizes \cite{binder2,majumder, majumder2,majumder4}. 
In the latter case data from smaller systems deviate from the
thermodynamic limit behavior earlier -- see the inset. In the case of disease dynamics the thermodynamic limit
(only in the sense of analogy that there is a crossover from this behavior to a saturation at a finite number)
behavior mostly comprises of the exponential period.

\begin{figure}[h!]
\centering
\includegraphics*[width=0.4\textwidth]{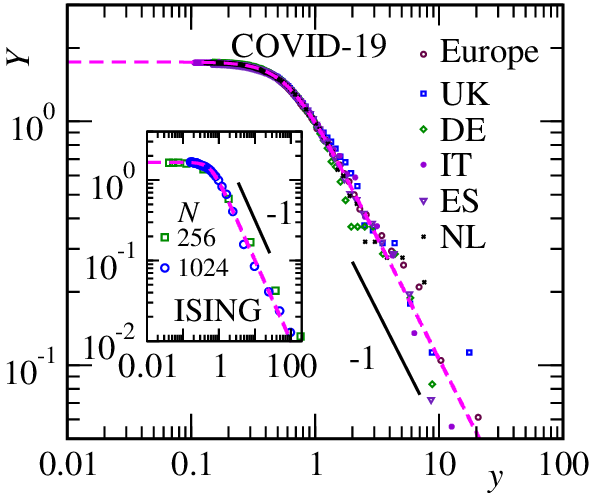}
\caption{\label{fig3}
Finite-size scaling exercise for COVID-19 fatalities in a group of regions:
$Y$ is plotted versus $y$, on a log-log scale. The solid line is a power-law with
	exponent $-1$. The dashed line represents the scaling function in Eq. (\ref{scl4}) with parameter
values $Y_0=1.06$, $b=0.23$, $\theta=2.93$, and $p=1.0$. Inset: Here the finite-size
	scaling function for COVID-19 is compared with the scaled Ising model data for two smallest studied system
	sizes for which finite-size limits have been reached within the time scales of our simulations. 
	For these systems, i.e., for $N=256$ and $1024$, the obtained values of $L_{\rm max}$, respectively, are
	$6.2$ and $12$. In this case we have obtained best
	data collapse for $\alpha=0.3$.
	Small discrepancy in $\alpha$, from the expected theoretical value of $1/3$,
	is due to the error in the domain size measurement, owing to thermal noise that has not been
	eliminated unlike some of our previous studies. Furthermore, for more accurate estimation of $\alpha$
	introduction of an instability time $t_0$ becomes necessary (see Ref. \cite{majumder}).}
\end{figure}

The main frame of Fig. \ref{fig3} shows the (finite-size) scaling plot for the 
disease dynamics. There data from all the above mentioned regions have 
been included. The collapse appears quite satisfactory, the quality being as good as that for those
obtained for phase transitions via computer simulations of simple model systems \cite{das0}. This confirms the similarity between the two types of problems.
The solid line in this figure represents a power-law with exponent $-1$ that corresponds to $p=1$, i.e., simple exponential
behavior at early time. The dashed line there is the analytical function in Eq. (\ref{scl4}).
The function parameters were obtained by fitting Eq. (\ref{scl4}) to the combined data set. For the values see the caption.
Near perfect collapse of data from many regions suggest robust universality in the
growth dynamics of the disease. Such universality, in spite of the differences in $m_f$, that,
for the considered places, lies in the range $[0.275,0.35]$
(Europe: $0.32$; DE: $0.275$; NL: $0.33$; IT: $0.325$; ES: $0.315$; UK: $0.35$), suggests that in the
post exponential periods the regions are maintaining similar discrepancies with each other as those during the
exponential periods. The other best collapse parameters ($N_{0f}$, $N_{df}$) for different regions, in the
same sequence as above, are ($2, 1500$), ($2, 255$), ($1, 151$), ($2, 545$), ($28, 1500$) and ($1, 465$).

In the inset of Fig. \ref{fig3} we have shown the analogous scaling plot by using data from the  
simulations of the Ising model. 
There it is clearly seen that the phase transition data
are very nicely described by the analytical function for the disease. This observation combines
the uniqueness in the disease spread with phase transition, except for the fact that
in one case to get the actual number one needs to go through an exponential transformation and
in the other case an algebraic transformation is necessary. It is worth mentioning here, even though we have
shown results for six regions, we believe that data from more countries will be described by the
same analytical function. In the phase transition scenario also, universality in finite-size scaling
behavior is broad -- this combines the considered Ising model with coarsening scenario in multi-component mixtures
and fluid systems \cite{das0,majumder2,majumder4}. For the latter, neither the scaling function was previously derived nor
such an agreement was obtained with a construction of any type.

The parameters that decide the universality in disease dynamics are $p$ and $\theta$. The others are simple multiplicative 
factors. Between $p$ and $\theta$, it has already been observed that $p=1$, i.e., early behavior is practically simple
exponential. Thus, $\theta$ is the quantity to look for. Note that for the Ising case, independent fitting to the combined 
scaling plot provides $\theta=2.29$, for $p=1$.  While this is only about $20$\% less than the COVID-19 value,
in a less restricted sense the universality can be
classified in accordance with power-law or exponential behavior of $f(y)$.

\begin{figure}[h!]
\centering
\includegraphics*[width=0.4\textwidth]{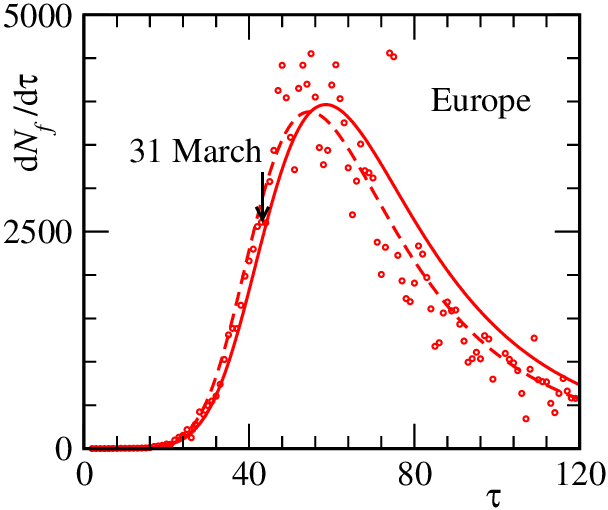}
\caption{\label{fig4}
Symbols represent the daily deaths, due to COVID-19,  in Europe for a period of 120 days, starting
        from $27$ January. The dashed line is obtained from the scaling function in Fig. \ref{fig3}.
        The continuous line is obtained by fitting only the data of Europe till $31$ March, by fixing
	$\theta$ at $3$, approximately the number for the scaling function in Fig. \ref{fig3}.}
\end{figure}

In Fig. \ref{fig4} we have demonstrated the predictive capability of the scaling model. There
we have compared the real data for daily new deaths in Europe with the analytical description. 
The connection between $Y$ and $N_f$ has already been mentioned above.
The dashed line in this figure represents analytical function obtained by fitting to the full combined
data set in Fig. \ref{fig3}. Obviously, the agreement is very good. For the prediction
purpose, however, we have fitted the data for Europe only till March 31, a date that is significantly earlier
than the appearance of the peak in this continent. Again the agreement is good. Certainly, then,
the model has strong predictive character. 

\begin{figure}[h!]
\centering
\includegraphics*[width=0.4\textwidth]{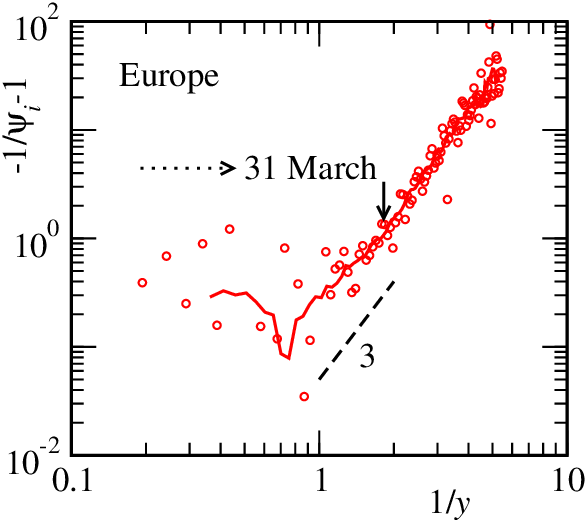}
\caption{\label{fig5}
Dependence of $\psi_i$, the instantaneous exponent, on $y$ is demonstrated,
on a double-log scale, for Europe. The dashed line represents $\theta=3.0$. The arrow-headed
solid line points to the date $31$ March. The arrow-headed dotted line shows the direction
        of increasing time. The continuous line represents running averaged plot of the
	original data that are shown using symbols.}
\end{figure}

In the fitting of the data for Europe till 31 March, we had fixed the value of $\theta$ at 
$3$, approximately the number that is obtained for $\theta$ by fitting the combined full data set in Fig. \ref{fig3}.
The justification is provided below.
Dependence of $\psi_i$ on $y$ is shown in Fig. \ref{fig5}. The number obtained by
fitting the combined data is consistent with the actual scenario. Furthermore, by $31$ March, a steady picture,
consistent with the above mentioned exponent, has already emerged. This tells about the reason behind 
such a good prediction. The overall story essentially conveys that a unique pattern for the disease
dynamics for many places emerges rather early. Once this is captured, action on SR and various facilities
becomes easier.

\section{Conclusion}

We have studied real data \cite{wiki} concerning fatality due to COVID-19 in a group 
of regions. A finite-size-like scaling model \cite{fisher3,book1,binder2,majumder,majumder2,majumder4,das0},
analogous to the one used for studies of phase transitions in small boxes, has been applied.
Analysis via this model suggests that the growth in fatality occurs exponentially fast at early time.
The model nicely combines this with the behavior at late time and suggests that there
exists strong universality \cite{skd_spread,jkb}. It has been shown that the coarsening behavior 
during phase transitions \cite{book1} follow the analogous scaling
pattern when studied in finite boxes \cite{binder2,majumder,majumder2,das0}.

An empirically derived finite-size scaling function \cite{skd_spread} describes
the disease growth as well as domain growth during phase transitions quite accurately.
The scaling function, and, thus, the model, has predictive power. This has been demonstrated
by analyzing data over a small early time window and comparing the analytical function for the
time dependence of $N_f$, thus obtained, with the future of the disease dynamics. The construction
of the function does not put restriction on the very early-time monotonicity. However, for a pure
exponential behavior it is necessary. This feature of the function may lead to errors if
the data are very noisy. However, conditions can be imposed to get rid of such errors.

Error free predictions indeed require high quality early time data. An
improvement by statistical averaging is not possible with real data, as far as the spread of a disease is concerned.
Nevertheless, models can be constructed \cite{sourav0} to replicate, in a statistical sense, the ground data, providing scope
for improvement by averaging over many such trials. We are interested in devising such combined
methods \cite{sourav1}



{\bf Acknowledgment:} 
The author acknowledges financial supports from the 
Department of Biotechnology, India (grant No. LSRET-JNC/SKD/4539); 
and Science and Engineering Research Board of Department of Science and Technology, 
India (grant No. MTR/2019/001585). He is thankful to F. M\"uller-Plathe for
important queries and to K. Dey for a help with the bibliography.

{\bf Data accessibility:} This article has no additional data.

{\bf Authors' Contributions:} S.K.D. designed the problem, carried out the work and wrote the manuscript.

The authors declares that there is no competing interest. 



~$^{*}$ das@jncasr.ac.in

\end{document}